\documentclass[
aps,
prresearch,
reprint,
superscriptaddress,
amsmath,amssymb,
floatfix,
]{revtex4-2}
\usepackage[utf8]{inputenc}
\usepackage[T1]{fontenc}
\usepackage{color}
\usepackage{graphicx}% Include figure files
\usepackage{dcolumn}% Align table columns on decimal point
\usepackage{bm}% bold math
\usepackage{gensymb}
\usepackage{orcidlink}
\begin{document}

% \preprint{APS/123-QED}

\title{Commissioning the Resonance Ionization Spectroscopy Experiment at FRIB}% Force line breaks with \\

\author{A.J.~Brinson\orcidlink{0000-0002-9551-5298}}
\email{brinson@mit.edu}
\affiliation{Laboratory for Nuclear Science, Massachusetts Institute of Technology, Cambridge, Massachusetts, USA.}

\author{B.J.~Rickey\orcidlink{0009-0004-5935-4505}}
\email{rickey@tamu.edu}
\altaffiliation{Present Address: The Cyclotron Institute at Texas A\&M University, College Station, Texas 77840, USA}
\affiliation{Facility for Rare Isotope Beams, Michigan State University, East Lansing, Michigan 48824, USA}
\affiliation{Department of Physics, Michigan State University, East Lansing, Michigan, USA}

\author{J. M. Allmond}
\affiliation{Physics Division, Oak Ridge National Laboratory, Oak Ridge, Tennessee 37831, USA}

\author{A. Dockery\orcidlink{0000-0002-3066-978X}}
\affiliation{Facility for Rare Isotope Beams, Michigan State University, East Lansing, Michigan 48824, USA}
\affiliation{Department of Physics, Michigan State University, East Lansing, Michigan, USA}

\author{A. Fernandez Chiu \orcidlink{0009-0007-6916-8986}}
\affiliation{Laboratory for Nuclear Science, Massachusetts Institute of Technology, Cambridge, Massachusetts, USA.}

\author{R.F.~Garcia~Ruiz}
\email{rgarciar@mit.edu}
\affiliation{Laboratory for Nuclear Science, Massachusetts Institute of Technology, Cambridge, Massachusetts, USA.}
\affiliation{Harvard-MIT Center for Ultracold Atoms, Cambridge, Massachusetts 02138, USA}

\author{T. J. Gray\orcidlink{0000-0003-3965-6130}}
\affiliation{Physics Division, Oak Ridge National Laboratory, Oak Ridge, Tennessee 37831, USA}
\affiliation{Department of Physics and Astronomy, University of Tennessee, Knoxville, Tennessee 37996, USA}

\author{J. Karthein\orcidlink{0000-0002-4306-9708}}
\altaffiliation{Present Address: Texas A\&M University, Cyclotron Institute / Dep. of Physics \& Astronomy, College Station, TX 77840, USA}
\affiliation{Laboratory for Nuclear Science, Massachusetts Institute of Technology, Cambridge, Massachusetts, USA.}

\author{T. T. King \orcidlink{0000-0003-0649-3275}}
\affiliation{Physics Division, Oak Ridge National Laboratory, Oak Ridge, Tennessee 37831, USA}

\author{K. Minamisono \orcidlink{0000-0003-2315-5032}}
\email{minamiso@frib.msu.edu}
\affiliation{Facility for Rare Isotope Beams, Michigan State University, East Lansing, Michigan 48824, USA}
\affiliation{Department of Physics, Michigan State University, East Lansing, Michigan, USA}

\author{A. Ortiz-Cortes}
\affiliation{Facility for Rare Isotope Beams, Michigan State University, East Lansing, Michigan 48824, USA}

\author{S. V. Pineda \orcidlink{0000-0003-1714-4628}}
\altaffiliation{Present Address: Université de Caen Normandie, ENSICAEN, CNRS/IN2P3, LPC Caen UMR6534, F-14000 Caen, France}
\affiliation{Facility for Rare Isotope Beams, Michigan State University, East Lansing, Michigan 48824, USA}
\affiliation{Department of Chemistry, Michigan State University, East Lansing, Michigan 48824, USA}

\author{B. C. Rasco \orcidlink{0000-0002-4061-1178}}
\affiliation{Physics Division, Oak Ridge National Laboratory, Oak Ridge, Tennessee 37831, USA}

\author{M. Reponen \orcidlink{0000-0003-1298-7431}}
\affiliation{Department of Physics, University of Jyväskylä, Accelerator laboratory, Jyväskylä, Finland}
%$mikael.h.t.reponen@jyu.fi$

\author{S. M. Udrescu \orcidlink{0000-0002-1989-576X}}
\altaffiliation{Present Address: Department of Physics and Astronomy, Johns Hopkins University, Baltimore, MD 21210, USA}
\affiliation{Laboratory for Nuclear Science, Massachusetts Institute of Technology, Cambridge, Massachusetts, USA.}

\author{A. R. Vernon}
\affiliation{Laboratory for Nuclear Science, Massachusetts Institute of Technology, Cambridge, Massachusetts, USA.}

\author{S. G. Wilkins\orcidlink{0000-0001-8897-7227}}
\altaffiliation{Present Address: Facility for Rare Isotope Beams, Michigan State University, East Lansing, USA}
\affiliation{Laboratory for Nuclear Science, Massachusetts Institute of Technology, Cambridge, Massachusetts, USA.}

%\collaboration{RISE Collaboration}%\noaffiliation

\date{\today}% It is always \today, today,

\begin{abstract}
This manuscript reports on the commissioning of the Resonance Ionization Spectroscopy Experiment (RISE) at the BECOLA facility at FRIB. The new instrument implements the collinear resonance ionization spectroscopy technique for sensitive measurements of isotope shifts and hyperfine structure of short-lived isotopes produced at FRIB. The existing BECOLA beamline was extended to integrate an electrostatic ion-beam bender and an ion detector at ultra-high vacuum. An injection-seeded Ti:Sapphire laser and a multi-harmonic pulsed Nd:YAG laser were installed to perform resonant excitation and selective ionization. Commissioning tests were performed to demonstrate the capabilities of the new instrument by measuring the hyperfine structure of stable $^{27}$Al produced in an offline ion source. The RISE instrument is ready and operational for future studies of short-lived isotopes at FRIB.

\end{abstract}

\maketitle

%\tableofcontents

\section{\label{sec:level1}Introduction}
The Facility for Rare Isotope Beams (FRIB) will dramatically expand access to exotic nuclei at the limits of stability~\cite{Ostroumov2024}, opening up new opportunities in our understanding of atomic nuclei and nuclear matter. In parallel with the development of radioactive beam facilities, it is essential to develop sensitive experimental techniques capable of measuring the properties of the most exotic nuclei, which typically have lifetimes shorter than a second and are produced at rates of only a few ions per second. Among the various spectroscopy techniques, collinear laser spectroscopy~\cite{Kaufman1976, Yan22, Ca-radii2019}---and in particular collinear resonance ionization spectroscopy~\cite{Sch91,Fla13,CRIS-NIM2013,Gro15,CRIS-NIM2020}---has been demonstrated to be a highly sensitive method for precisely measuring the electromagnetic properties of ground and long-lived isomeric states of exotic nuclei~\cite{deGroote2020MeasurementIsotopes,K-radii2021,Ver25}. Recently, 
collinear laser spectroscopy has been used to infer an important parameter in the equation of state for dense nuclear matter~\cite{Brown2020ImplicationsState, Pineda2021ChargeRadii, Konig2024NuclearIsotopes}, and 
collinear resonance ionization spectroscopy has even been used to measure radioactive molecules~\cite{GarciaRuiz2020}, 
providing isotopologue-shift~\cite{Udr21} and hyperfine-structure measurements in these systems~\cite{Wilkins2025ObservationMolecule}, and guiding the next generation of fundamental symmetry searches~\cite{Udrescu2024}.

By overlapping lasers with a fast ion beam accelerated typically to an energy of a few tens of keV in a collinear geometry, Doppler-broadening-suppressed measurements of electronic transitions can be performed due to the kinematical compression~\cite{Kaufman1976}. These measurements are sufficiently sensitive to resolve isotope shifts and hyperfine structures, and allow for the extraction of changes in nuclear charge radii, as well as determinations of nuclear spins, magnetic dipole moments, and electric quadrupole moments across isotopic chains~\cite{Ver22,Karthein2024}.

Here, we present a new collinear laser spectroscopy instrument, the Resonance Ionization Spectroscopy Experiment (RISE), installed at the BEam COoler and LAser spectroscopy (BECOLA) facility~\cite{Minamisono2013CommissioningNSCL,
Rossi2014ABeams} %, Minamisono2016ChargeFragmentation
at FRIB. Details of the instrument are presented in Sec. \ref{sec:layout}. Results from the commissioning tests of the new instrument using stable $^{27}\text{Al}$ are reported in Sec. \ref{sec:comissioning}. Finally, conclusions and perspectives for new experiments are discussed in Sec. \ref{sec:conclusion}. 

\section{\label{sec:layout}Experimental layout}
\begin{figure*}[ht]
\includegraphics[width=1\textwidth]{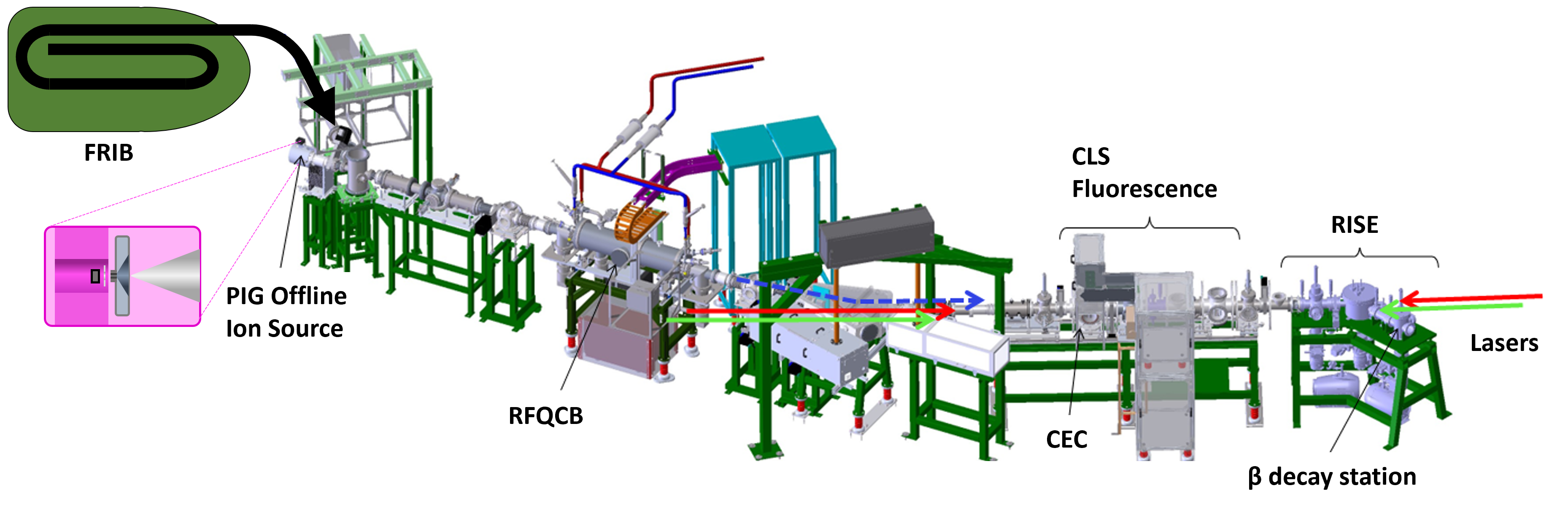}
\caption{\label{fig:rise} \textbf{Layout of the BECOLA facility at FRIB.} 
%Top: Rare isotopes are produced by the FRIB linear accelerator via projectile-fragmentation reactions with heavy ions on a thin foil target, which are then filtered by the ARIS in-flight separator to select the isotope of interest, and thermalized in the gas stopper and extracted at low energy (tyically 30 keV), before being sent to BECOLA. Bottom: 
In the BECOLA facility, radioactive (via FRIB) or stable (via PIG source) ion beams are injected into the RFQCB, extracted at an energy of $\sim$30~keV, and neutralized in-flight. Anti/collinear laser spectroscopy is then performed either via fluorescence detection, or via single ion counting with the new RISE beamline extension. 
%A removable mirror located at the bending chamber near the RFQCB can be used to reflect the spectroscopy laser by 180$\degree$, enabling collinear spectroscopy.
}
\end{figure*}

A schematic of the BECOLA facility~\cite{Minamisono2013CommissioningNSCL, Rossi2014ABeams} is shown in FIG. \ref{fig:rise}. Prior to the RISE extension, the facility consisted of four main components: \textit{a.} ion source; \textit{b.} radiofrequency quadrupole cooler buncher (RFQCB); \textit{c.} charge-exchange cell; and \textit{d.} fluorescence detection unit. An overview of the pre-existing BECOLA components is given first, followed by a detailed discussion of the RISE instrument.

%The BECOLA facility is a collinear laser spectroscopy system used for studying nuclear properties of exotic isotopes at FRIB (formerly NSCL). 
%The BECOLA experimental procedure can be broken down into 5 primary steps:
\subsection{\label{sub:overview} BECOLA overview}
\paragraph*{Online rare isotope production}
During an experiment on rare isotopes, exotic nuclei are produced at FRIB via in-flight projectile-fragmentation reactions of heavy ions on a thin production target. An isotope of interest is then selected from many reaction products via the ARIS fragment separator~\cite{Hausmann2013DesignFRIB} and transported to the gas-stopping system. The ion beam is slowed through collisions with degraders and ultimately stopped in a helium-filled gas stopping cell~\cite{Vil23}. Once thermalized, the ions are extracted at an energy of typically 30 keV and transported to the BECOLA facility.

\paragraph{Ion source}
A dedicated offline ion source to produce stable isotopes is installed at the BECOLA facility. A Penning ionization gauge (PIG) ion source (a plasma discharge sputtering ion source) is used~\cite{Nouri2010ASpread, Ryder2015PopulationVapor} to produce stable isotopes of metallic and gaseous elements. The availability of a stable offline source is essential for conducting offline tests to develop and optimize the experimental design (e.g. transition and detection scheme development)~\cite{Dockery2023HyperfineII} in preparation for experiments on radioactive isotopes, which are generally produced in limited quantities. The offline ion source is also critical during online experiments, to calibrate the ion beam energy~\cite{Konig2021BeamSpectroscopy}, and to monitor the stability of the laser spectroscopy system over the course of a typically several-day measurement period.

\paragraph{Radiofrequency quadrupole cooler buncher}
Ion beams from either the FRIB gas stopping cells or the BECOLA PIG ion source are injected into the RFQCB~\cite{Barquest2017RFQIsotopes}, where they are collisionally cooled by 300~K helium buffer gas.
The RFQCB has separate sections for cooling and bunching, with the bunching section being maintained at $\sim$100 times smaller vacuum (typically $10^{-5}$~Torr) via differential pumping. Quadrupole rod electrodes are driven at radiofrequency to provide radial confinement, while segmented DC electrodes provide potential to confine the ions longitudinally and to transfer them between regions. By switching off the potential applied to the last DC electrode, the trapped ions are released as a bunch and accelerated to high velocity, due to the potential difference maintained between the RFQCB and the remainder of the beamline.

%This cooling
% serves to reduce the transverse and longitudinal emittances of the ions, which 
%improves the emittance and transport efficiency through the remainder of the beamline.
%, and reduces the Doppler broadening of the resonant spectra.
% While there is a strong dependence on the RF setting used in the trap, emittances as low as as 2 $\pi$mm mrad have been measured.

%A comprehensive description of collinear laser spectroscopy is given in reference \cite{Kaufman1976}. Following this acceleration, the ions pass through a 30$\degree$ electrostatic bender, allowing for spatial overlap with lasers later on.
% , without requiring laser light to pass through the RFQCB or any components further upstream. A set of Faraday cups are positioned throughout the beamline to diagnose transport efficiency, and a series of electrostatic lenses and deflectors are used to optimize this transport.
\paragraph{Charge-exchange cell}
In the typical case where the most desirable spectroscopy transition is in the neutral species, it is necessary to neutralize the fast moving ion bunch prior to the interaction with the laser light. Thus, neutralization can be accomplished in-flight by passing the ion bunch through an alkali (e.g. sodium) vapor produced in a charge-exchange cell (CEC)~\cite{Klose2012TestsSpectroscopy}. As the ions pass through the cell, they interact with the alkali vapor, inducing charge-exchange reactions that produce a high neutral yield in the outgoing bunch.
A pair of deflector electrodes is located at the exit of the CEC to remove any remaining ions from the bunch.
A scanning voltage is applied to the CEC via a Matsusada DC amplifier (AMP-10B10-02), driven by a small ramping voltage, which allows the ion bunch velocity (and thus the subsequently neutralized atom bunch velocity) to be modulated. This velocity modulation brings the Doppler-shifted laser frequency into resonance and allows BECOLA to scan through resonant structures without needing to change the laser frequency in the lab frame. A detailed discussion of the BECOLA CEC can be found in Ref.~\cite{Klose2012TestsSpectroscopy}.

\paragraph{Fluorescence detection} 
A single laser is overlapped with the (neutral or ion) bunch trajectory in the photon detection region \cite{Maa2020ASpectroscopy} of the BECOLA beamline. The laser can be input coupled at either end of the beamline, enabling both collinear and anticollinear spectroscopy to be performed. When the bunch is Doppler shifted onto resonance with the laser frequency, the atoms are excited and subsequently emit spontaneous fluorescence. This fluorescent emission is collected by an ellipsoidal reflector and detected by a Photomultiplier tube (PMT). By correlating the PMT count rate with the total kinetic energy of the bunch, a frequency spectrum of the atomic structure is obtained.

The sensitivity of fluorescence detection is significantly limited by the background from scattered laser light and the relatively low efficiency of photon collection and detection. Moreover, this method is best suited only for short-lived atomic transitions, with lifetimes typically on the order of tens of nanoseconds or less, due to the limited geometry of the ellipsoidal reflector ($\sim$5~cm along the atom beam).
To overcome these limitations, RISE offers a complementary detection scheme based on direct particle detection, which typically enhances detection efficiency and significantly reduces background noise compared to fluorescence detection.

\subsection{\label{sub:res}Resonance ionization in RISE}
In resonance ionization spectroscopy, a chain of two or more transitions is driven within the atom, with the final transition yielding ionization. The resulting ion is then deposited onto an ion detector, and by correlating the detector count rate with the kinetic energy of the atom beam bunch, a hyperfine spectrum can be constructed. While fluorescence detection is limited by the solid angle for photon collection and the quantum efficiency of photomultiplier tubes (typically below 25\%), ions can be electrostatically steered, and the efficiency of ion counting exceeds 80\%.
Furthermore, the background rate for ion counting (in most cases dominated by collisional ionization with residual gas,
and by non-resonant ionization of long-lived states populated during the charge-exchange process) depends linearly on the ion beam intensity, which is typically very small for radioactive beams. In contrast, the background in fluorescence detection is dominated by scattered laser light, which is constant for a given laser power and can be as much as one thousand times larger than the background from ion detection.

\begin{figure*}[ht]
\includegraphics[width=1\textwidth]{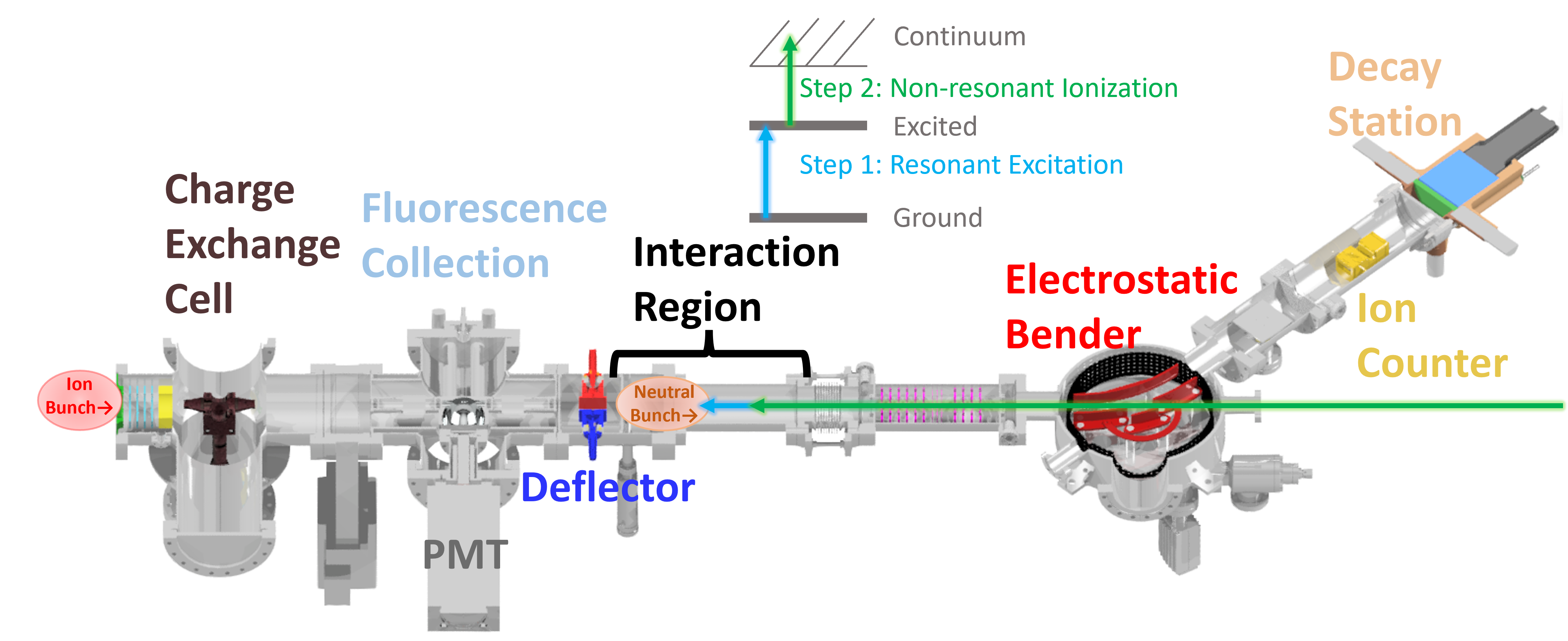}
\caption{\label{fig:rise_layout} \textbf{Layout of the RISE extension to the BECOLA beamline.} Bunches of cooled ions from the RFQCB are sent to the CEC for neutralization. For fluorescence experiments, a laser is overlapped with the neutral bunch directly after the CEC, and the resulting fluorescence is collected and detected by a set of PMTs. For resonance ionization experiments, the non-neutral bunch fraction is filtered out by the deflector, while the neutral fraction is overlapped with lasers in the interaction region, causing resonant excitation, followed by selective ionization.
%In place of laser ionization, field ionization plates can also be used.
The reionized bunch is then steered onto a retractable MagneTOF single ion counter after a 30$\degree$ electrostatic bender. A $\beta$-decay station is located at the end of the beamline, for beam characterization during experiments with rare isotope beams and for laser-assisted nuclear decay spectroscopy \cite{Lynch2016Combined198}.}
\end{figure*}

%\subsection{Resonance Ionization Spectroscopy Experiment}
A schematic layout of the RISE instrument is shown in FIG.~\ref{fig:rise_layout}. Following the neutralization process, a deflector is used to remove any remaining ions. The resulting neutral atomic beam is then overlapped with multiple laser beams along an interaction region maintained at ultra-high vacuum, typically below $5\times10^{-9}$~mbar with the CEC operating. First, a narrow linewidth laser is used to resonantly excite an atomic transition from the ground or a metastable state to an excited state. This laser is input through a gap in a 30$\degree$ electrostatic bender. For collinear spectroscopy, a removable mirror, which is located on an optical breadboard next to the bending chamber nearest the RFQCB, can be used to reflect the beam by 180$\degree$.
% , from which the hyperfine structure and transition centroid can be deduced.
Selective ionization can then be carried out from the excited state. To further suppress the background due to ionization of inadvertently populated excited states, a sequence of multiple resonant excitations may be employed, enabling a final ionization step with lower energy and thus higher selectivity. Finally, ions that were resonantly excited and selectively ionized are steered through a 30$\degree$ electrostatic bender and onto the ion counting unit. Ion counting is performed by a MagneTOF particle detector, a fast electron-multiplier–based time-of-flight detector that provides sub-nanosecond timing resolution, ion detection efficiencies larger than 80\%, and a large dynamic range that remains linear for ion bursts exceeding $3\times10^5$ events. The intrinsic background of the MagneTOF is very low, with typical dark count rates below 20 counts per minute at operating voltages around 3 kV, making the detector well suited for low-count-rate measurements. When ions are impinged onto the MagneTOF, the resulting current pulse is fed into a shunt resistor to produce a voltage signal, which is amplified, discriminated, digitized, and recorded with a ns-resolution timestamp by a custom field-programmable-gate-array-based time-resolved scaler \cite{Rossi2014ABeams}.

\subsection{Laser systems}
\begin{figure}[h]
\includegraphics[width=0.5\textwidth]{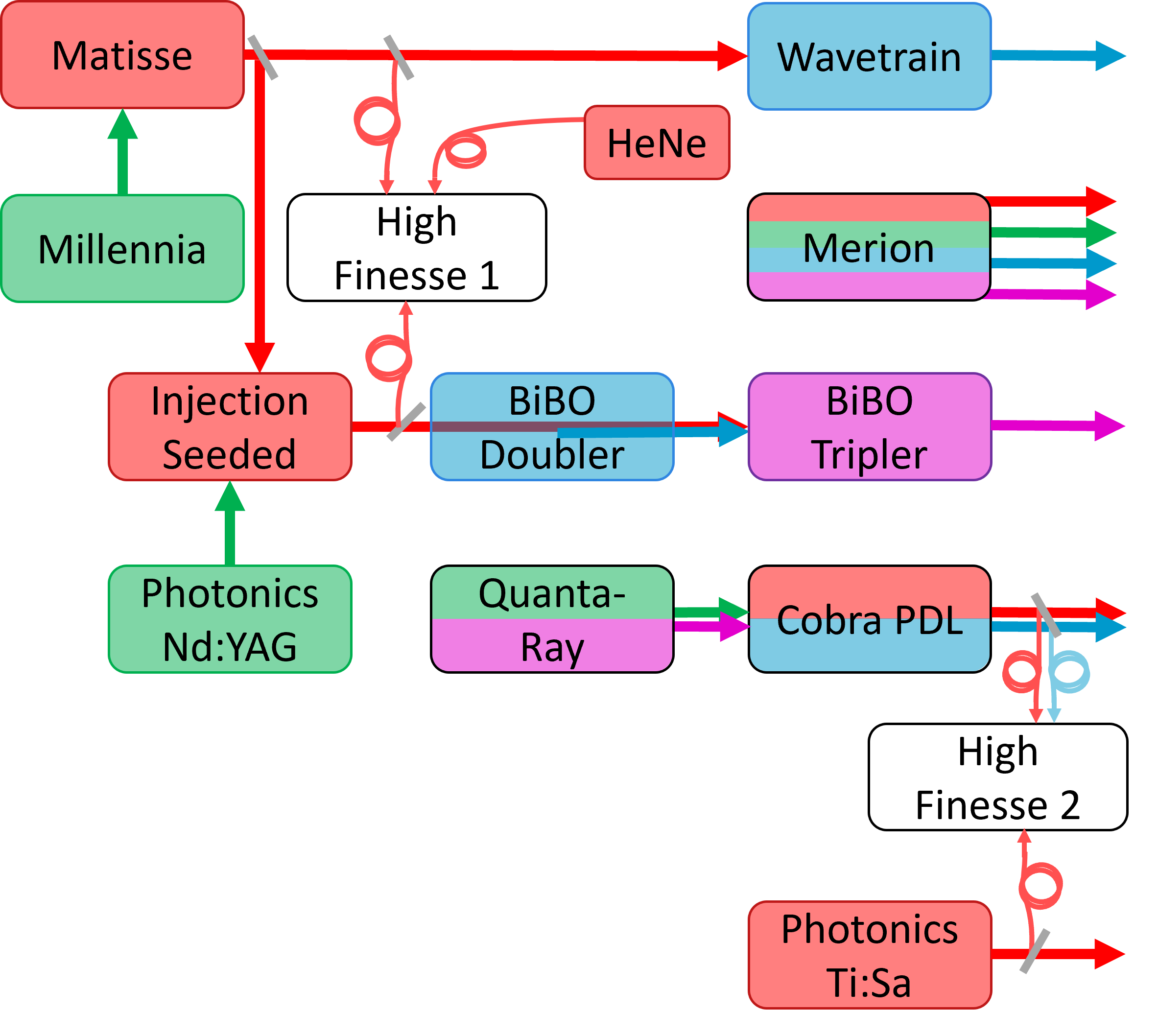}
\caption{\label{fig:laser} \textbf{Schematic diagram of RISE/BECOLA laser systems.} White boxes indicate wavemeters, while red ($650-1000$~nm), green($532$~nm), blue ($350-500$~nm), and violet ($\lesssim350$~nm) boxes indicate lasers at various wavelengths. Outgoing arrows indicate coupling into the beamline. The helium-neon (HeNe) laser is used to keep the HighFinesse WSU30 (Wavemeter 1) calibrated.
}
\end{figure}

A schematic diagram of the laser systems available at RISE/BECOLA is shown in FIG.~\ref{fig:laser}. 

\subsubsection{Resonant excitation}
A continuous-wave (CW) titanium-sapphire (Ti:Sa) laser (Spectra-Physics Matisse), pumped by a frequency-doubled neodymium-doped yttrium aluminum garnet (Nd:YAG) laser (Spectra Physics Millenia), is used to produce laser light between 700-1000 nm, the frequency of which can be locked to a precision of 1~MHz, and recorded with a HighFinesse WSU30 wavemeter. For consistent measurement, the wavemeter is regularly calibrated with a frequency-stabilized helium-neon laser.  The Matisse output is then fiber coupled to be used as the seed of an injection-seeded Ti:Sa cavity~\cite{Laser-injec, Reponen2018TowardsPALIS-facility}. The latter is pumped by a 532~nm Nd:YAG laser (Photonics TU-H), generating pulses of 30-50~ns at a repetition rate of 10~kHz. The output of the injection-seeded cavity is then narrowband pulses of about 20 MHz linewidth, with energies on the order of 200~$\mu$J, locked to the frequency of the seed laser. These relatively high-energy pulses can then be focused through one or more $\beta$-BaB$_2$O$_4$ (BBO) and/or BiB$_3$O$_6$ (BiBO) non-linear crystals, to produce doubled, tripled, or quadrupled frequencies in a single-pass conversion unit. This laser is typically used to drive the "spectroscopy" transition, from which hyperfine spectra and isotope shifts are deduced.

When driving a chain of transitions, it is often desirable to use broader linewidth lasers for all steps except the spectroscopy transition. The broad linewidth covers the hyperfine splitting and isotope shift, and even the variation in Doppler shift caused by the mass difference when measuring different isotopes. A pulsed dye laser (PDL) or broadband Ti:Sa can be used for such multi-resonance schemes, allowing for higher selectivity ionization, as well as spectroscopy of Rydberg and autoionizing states. A PDL (Sirah Cobra Stretch), pumped by a frequency-doubled (532~nm) or -tripled (355~nm) Nd:YAG laser (Spectra Physics Quanta-Ray), can be filled with a wide range of dyes, enabling the generation of pulsed laser light spanning from 360 - 800 nm, with linewidths on the order of 10~GHz. While the PDL is too broad to resolve most hyperfine structures or the field-shift components of isotope shifts in the light region of the nuclear chart, it is ideal for driving additional transitions in schemes involving multiple resonant steps prior to the ionization step, as performed in Ref. \cite{Udrescu2024}.

Additionally, a collection of three broadband grating-tunable pulsed Ti:Sa lasers (Photonics TU-H \cite{Liu2016AORNL}) can be used to reach wavelengths in the range of 720-960 nm, with a typical linewidth of $\sim$3~GHz.
% This is favorable for driving transitions to higher-lying (e.g. Rydberg or auto-ionizing) states.
These lasers are also configurable for second, third, and fourth harmonic generation.

\subsubsection{Selective ionization}
The principle behind selective ionization is to drive ionization via a mechanism that is only sufficiently energetic to ionize atoms which were already resonantly excited during previous steps.

A Quantel Merion MW 7-100 laser system is used to drive the non-resonance ionization. The Merion is a 100~Hz pulsed Nd:YAG system, with pulse width $<$ 10~ns. The Merion's modular design allows for switching between fundamental~(1064~nm), doubled~(532~nm), tripled~(355~nm), or quadrupled~(266~nm) outputs, with pulse energies up to 300, 160, 90, or 30~mJ, respectively.
%\textcolor{red}{TODO: cite https://www.quantel-laser.com?}
While the Merion's third and fourth harmonics have higher photon energy to reach the ionization threshold from lower-lying states, the doubled and fundamental outputs can be more desirable for background suppression, depending on the excitation scheme employed and the ionization energy of the species being studied.

\subsection{Decay station}
\begin{figure}[h]
\includegraphics[width=0.5\textwidth]{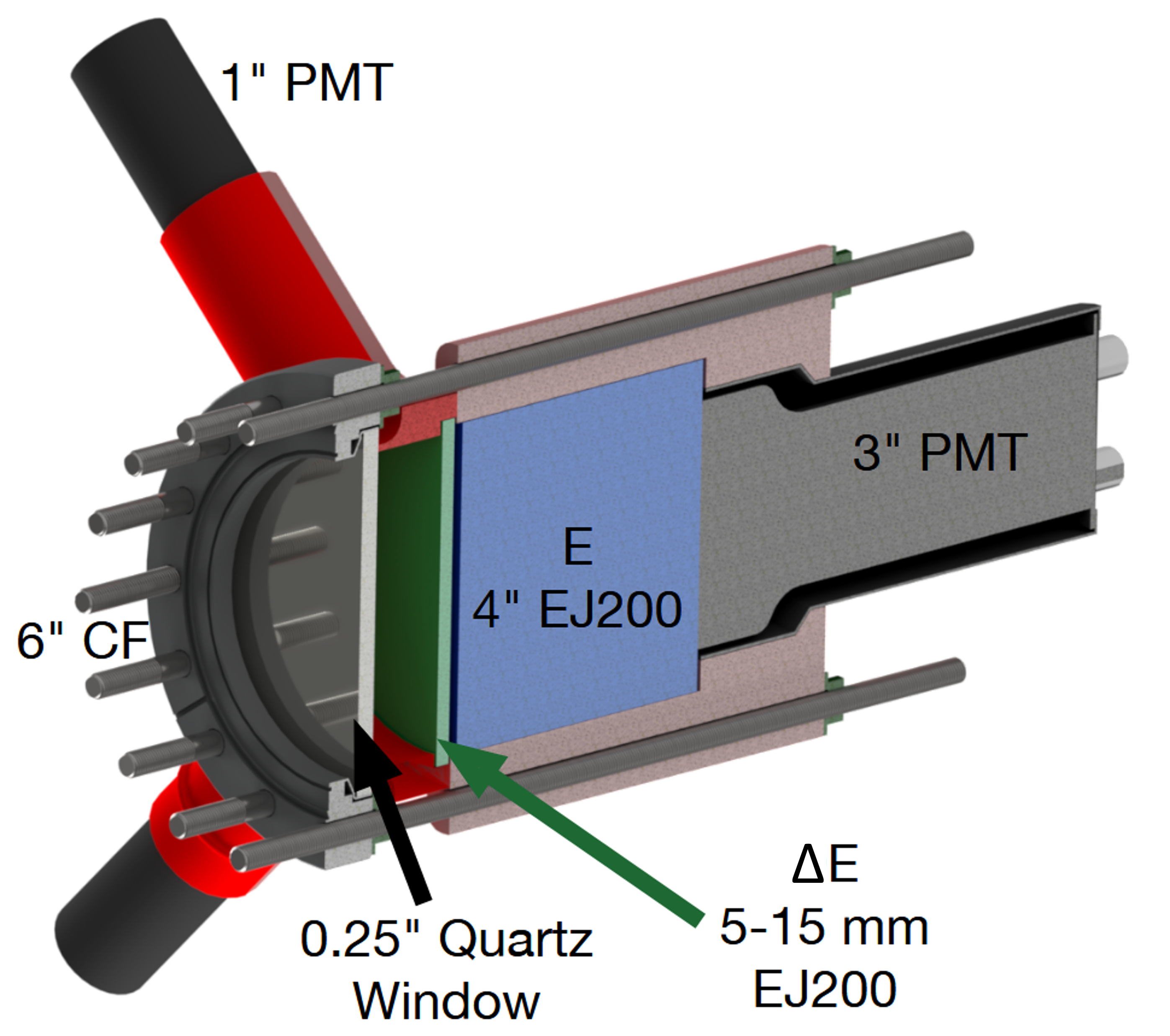} 
\caption{\label{decaystation} \textbf{$\beta$ decay station} Schematic of the $\beta$-decay station at the end of RISE, 
consisting of a $\Delta$E-E $\beta$ telescope mounted on a quartz viewport. The setup provides $\beta$ Q-value selectivity while suppressing muon and $\gamma$-ray background. The $\Delta$E detector is configurable for different decay energies. See text for more details.}
\end{figure}

The RISE extension terminates with a decay station, currently consisting of a $\Delta$E-E $\beta$-telescope attached to a 6-inch CF viewport.
The quartz viewport is 0.25 inches thick with a 1 micron silver coating evaporated on the vacuum side. The silver coating has an approximately 1-mm gap around the edge for electrical isolation, enabling beam current readout. The $\Delta$E and E detectors are made from EJ-200 plastic scintillator. The $\Delta$E detector is adjustable to match the experimental decay energies, and has enhanced specular reflectors on the front and back faces for light isolation. It is comprised of 4.5-inch point-to-point octagons with 5-, 10-, and 15-mm thick options, and is read out on the edges by four 1-inch PMTs. The E detector is a 4-inch diameter and 4-inch long cylinder backed by a 3-inch PMT. The geometric efficiency is roughly 22\% to the front face of the $\Delta$E and 3\% to the back face of the E. A drawing of the telescope can be seen in FIG.~\ref{decaystation}.

The telescope provides $\beta$ Q-value selectivity while suppressing muon and gamma-ray background. The most exotic isotope in the beam will have the largest $Q$-value. $\beta$-tagging has been demonstrated to be a highly sensitive detection scheme when combined with collinear resonance ionization spectroscopy, as it is insensitive to stable contaminants and long-lived isotopes sometimes present in radioactive beams \cite{K-radii2021, Powel2022Ground40}. At FRIB, this can be due to the small rare-gas contamination in the He stopping gas, and to the molecules produced during the gas-stopping process.

In addition to serving as an unstable beam diagnostic and a detector for atomic spectroscopy, the inclusion of a $\beta$-decay station in a resonance ionization beamline enables the possibility of carrying out nuclear decay spectroscopy with extremely high initial state purity \cite{Lynch2016Combined198}. By selectively ionizing a single atomic state with MHz-level resolution, RISE can isolate, for example, an isomer from the ground state, delivering only the desired nuclear state to the decay station.

\section{\label{sec:comissioning}Commissioning tests}

\begin{figure}[h]
\includegraphics[width=0.5\textwidth]{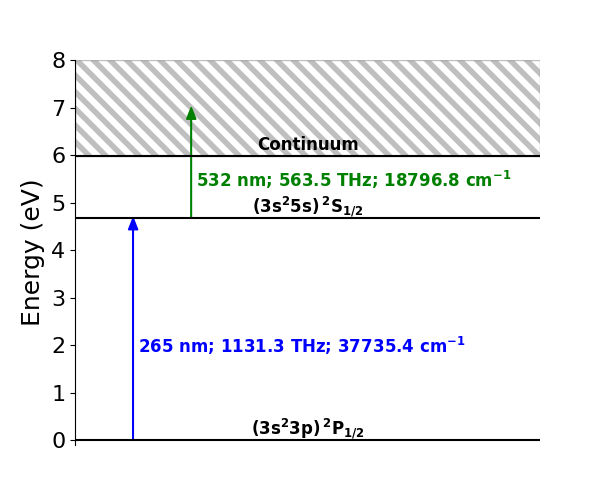}
\caption{\label{fig:levels} \textbf{Simplified level diagram for resonance ionization spectroscopy of Al.} 
The horizontal bars denote the energy levels of aluminum which were addressed in the commissioning tests, and are labeled by their electronic configurations and term symbols. The blue arrow indicates the single photon energy of the resonant excitation laser, while the green arrow corresponds to the photon energy for the selective ionization laser.
}
\end{figure}

\subsection{Procedure}
RISE was commissioned using a two-step resonance ionization scheme to investigate the $(3s^23p)~^2P_{1/2}\rightarrow (3s^25s)~^2S_{1/2}$ transition at 265~nm in $^{27}$Al, hereafter referred to as $^2P_{1/2}$$\rightarrow$$ ^2S_{1/2}$, for brevity. Aluminum ions were produced using the PIG ion source \cite{Ryder2015PopulationVapor}, injected into the RFQCB, cooled with 300~K He buffer gas, bunched, and ejected into the beamline at a duty cycle of 100~Hz to synchronize with the repetition rate of the ionization laser. The RFQCB potential was stabilized by a proportional feedback system, to be $29,916.8(4)$~V\cite{Konig2024HighApplications}
relative to the remainder of the beamline. The bunched ions were then steered onto the collinear laser spectroscopy beam line, velocity-modulated at the CEC and neutralized in-flight with sodium vapor. Resonant excitation light was provided by the injection-seeded Ti:Sa, locked to a frequency of 377.214828(10)~THz (collinear), or 376.052850(10)~THz (anticollinear), and subsequently frequency-tripled by a combination of BiBO and BBO crystals to produce pulses of 1131.644484(30)~THz (collinear), or 1128.155559(30)~THz (anticollinear) laser light. For selective ionization, the Merion MW 7-100 was used in the frequency-doubled (532~nm) configuration, with a pulse rate of 100~Hz and an average pulse energy of $\sim$34~mJ. The excitation (alternating between anti/collinear) and ionization (anticollinear) lasers were overlapped with the Al bunch, with the timings tuned so that the ionizing pulse arrived 40(20)~ns after the excitation pulse, as measured with a Thorlabs DET10A photodetector. This relative pulse timing was selected to limit the effect of AC Stark shift induced by the ionization laser, without sacrificing significant efficiency from spontaneous emission between the two pulses. The reionized Al was then steered onto the ion detector and recorded as timestamps relative to the trigger signal to release the ion bunch from the RFQCB. The ions were counted as a function of the scanning voltage applied to the CEC to be converted to the rest-frame laser frequency via Doppler shift.

\subsection{Spectral features}
\begin{figure}[h]
\includegraphics[width=0.5\textwidth]{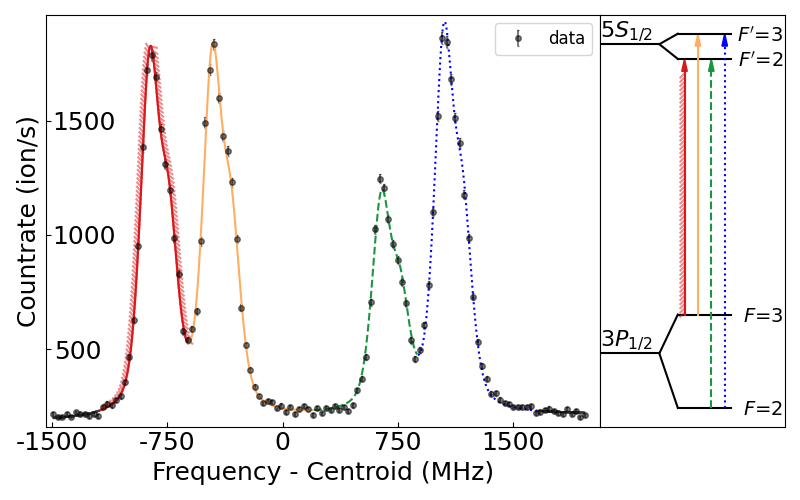}
\caption{\label{fig:hfs} \textbf{Example hyperfine spectrum for $^{27}$Al.} The four distinct peaks are due to the hyperfine splitting of the $^{2}P_{1/2}$ and $^{2}S_{1/2}$ states. The colors and types of fitted line indicate the corresponding transitions shown in the right-hand side of the figure. The individual hyperfine resonances have a common asymmetric line shape due to the energy exchange during the charge exchange process, manifesting as right-shifted side peaks for the anticollinear spectrum shown here (see text for more details). The error bars are of a similar size or smaller than the markers.}
\end{figure}

An example spectrum is shown in FIG.~\ref{fig:hfs}. Since $^{27}$Al has a nuclear spin ($I=\frac{5}{2}$), and the $^2P_{1/2}$ and $^2S_{1/2}$ states both have total electronic angular momentum $J=\frac{1}{2}$, the ground and excited states are each split into two levels, corresponding to $F$ = 2 or 3, where $F$ is the total angular momentum, $\bm{F}=\bm{I}+\bm{J}$. This gives rise to four distinct hyperfine transitions, with resonant frequencies $\nu_{FF'}$ given by:
\begin{equation}
    \nu_{FF'} = \nu_0 + \frac{1}{2}\left(A_{^2S_{1/2}}K_{F'IJ'}-A_{^2P_{1/2}}K_{FIJ}\right),
    \label{eqn:transitionFreqs}
\end{equation}
where $F'$ and $J'$ denote the excited state momenta, $K_{FIJ}\equiv F(F+1)-I(I+1)-J(J+1)$ is a measure of the alignment between the vectors $\bm{I}$ and $\bm{J}$, $A_X$ is the hyperfine constant associated with state $X$, and $\nu_0$ is the centroid of the transition.

The recorded resonant structures are well described by a pseudo-Voigt profile,
%, since the linewidth is due to a combination of homogenous (natural linewidth, power broadening) and inhomogenous (Doppler spread) factors
 with the addition of one or more side peaks, due to energy exchange between the Al beam and sodium during the neutralization process \cite{Bendali1986Na+-NaSpectroscopy, Klose2012TestsSpectroscopy, Dockery2025SpectralExchange}. Since these exchanges reduce the beam energy, they affect the Doppler shift in opposing ways for (anti-)collinear laser propagation, and so the side peaks appear on the (right)left side of the primary peak. These right and left side peaks can be seen by comparing the collinear and anticollinear spectra shown in FIG.~\ref{fig:bea}.

Since the mechanisms that produce these side peaks are independent of the hyperfine structure, a given class of side peaks can be characterized across all hyperfine transitions with two parameters: the frequency shift from the primary hyperfine resonance, and the proportion relative to the primary peak amplitude.
%These side peaks can also be constrained by simulating the neutralization process in the CEC \cite{Dockery2025SpectralExchange}.
%, further reducing the free parameters required to fit a spectrum.

By fitting these spectra, the centroid of the electronic transition, as well as the hyperfine constants of the states involved, can be deduced.
%The results obtained are summarized in TABLE \ref{tab:Constants Table}.
During the fitting procedure, the amplitudes of each peak are left free, while the lineshapes are constrained to share the same width and side peak parameters, and the peak centers are constrained by equation  (\ref{eqn:transitionFreqs}).

\subsection{\label{sub:bec}Beam energy correction}
\begin{figure}[h]
\includegraphics[width=0.5\textwidth]{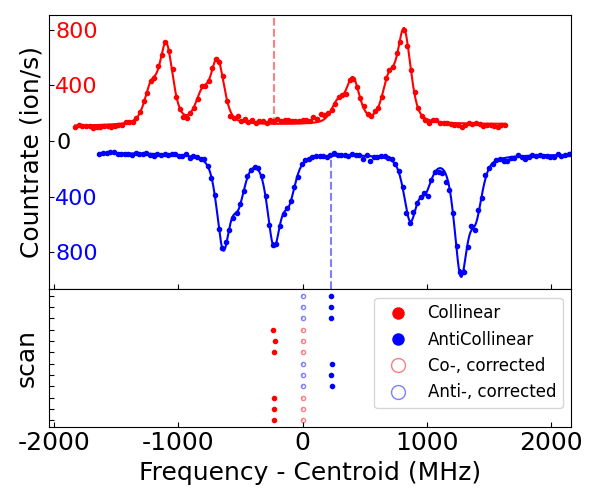}
\caption{\label{fig:bea} \textbf{Beam-energy correction.} Top: Example spectra for collinear (red) and anticollinear (blue) spectroscopy. Bottom: After correcting the beam energy, the deduced centroids from collinear and anticollinear measurements are consistent.}
\end{figure}

Since the bunch is moving relative to the lab frame, a proper account of the total beam energy is necessary for a correct application of the Doppler-shift transformation and subsequent extraction of an absolute transition frequency. An accurate voltage divider (Caddock film resistor) and a voltage divider system \cite{Konig2024HighApplications} were used to measure the scanning voltage applied to the CEC and the high voltage that provided the main beam energy, respectively. However, the actual beam energy can be shifted from the total applied voltages due to the trapping potential in the RFQCB, the atomic charge exchange and possible resistive voltage drop. In order to measure the actual beam energy precisely, a collinear/anticollinear measurement is performed, where the same transition is recorded twice: once with the excitation laser propagating anticollinearly to the bunched atomic beam, and then again with the laser propagating collinearly. By determining the Doppler-shifted centroid frequencies for each propagation direction, a Doppler-free centroid (absolute transition frequency) can be calculated, as detailed in Ref. \cite{Konig2021BeamSpectroscopy}.
An estimate for the rest frame centroid of the $^{2}P_{1/2}$~$\rightarrow$~$^{2}S_{1/2}$ transition is reported in TABLE \ref{tab:Constants Table}. This estimate is based on 6 collinear and 6 anticollinear measurements, constituting 1500 seconds of total integration time. This collinear/anticollinear analysis is also summarized in FIG.~\ref{fig:bea}.

Once an energy corrected $\nu_0$ is determined, it can in turn be used in combination with any collinear or anticollinear Doppler shifted centroid measurement, $\nu_{c/a}$, to reconstruct the total beam energy of a recorded spectrum, $E_{\rm kin}$, as:
\begin{equation}
    E_{\rm kin} = \frac{mc^2}{2}\frac{\left(\nu_0-\nu_{c/a}\right)^2}{
    \nu_0\nu_{c/a}
    },\label{absolute_beam_energy}
\end{equation}
where $m$ is the mass of the atom and $c$ is the speed of light. This yields a method for calibrating the beam energy throughout the course of an experiment.

\subsection{\label{sub:stability}Stability}
\begin{figure}[h]
\includegraphics[width=0.5\textwidth]{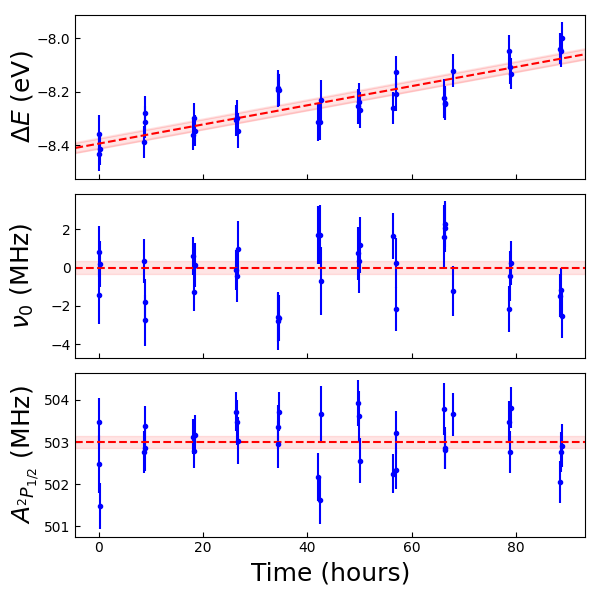} 
\caption{\label{fig:stability} \textbf{Measurement stability.}
Summarized results from 34 anticollinear spectroscopy measurements on the $^2P_{1/2}\rightarrow ^2S_{1/2}$ transition, taken over the course of 90 hours.
Top: Deviation of the absolute beam energy from the preset beam energy evaluated using the  anti/collinear analysis with Eq. (\ref{absolute_beam_energy}). A small systematic drift with a magnitude of 0.3~eV relative to $\sim$30 keV total energy is observed.
Middle: Remaining variation in centroid frequencies after applying a linear calibration to the beam energies as a function of time.
Bottom: Variation in the ground state hyperfine $A_{^2P_{1/2}}$ coefficient over the course of the commissioning tests presented here.}
\end{figure}

Since the production yields for unstable isotopes can be as low as a few ions per second, successful measurements can require more than a day of integration time to collect sufficient statistics on a single isotope. As such, for the extraction of relative measurements from absolute quantities---such as isotope shifts from differences in absolute transition frequencies---it is important that RISE provides a stable measurement environment over long periods of time. FIG.~\ref{fig:stability} shows the results of repeated measurements of the same $^2P_{1/2}\rightarrow ^2S_{1/2}$ transition over the course of 90 hours. While a non-negligible drift in the beam energy (0.0036(3) eV/hr) was observed over the course of these measurements, application of a linear beam energy correction in time was sufficient to bring the centroid estimates into agreement with the centroid derived from a previously performed anti/collinear analysis. The beam-energy-calibrated spectra yielded residual standard deviations of $\pm1.5$~MHz in the transition centroid, and $\pm0.6$~MHz in the $A_{^2P_{1/2}}$ hyperfine constant. This level of precision will be sufficient to measure many properties of interest across the nuclear chart.

An estimate for $A_{^2P_{1/2}}$ is given in TABLE \ref{tab:Constants Table}, based on a weighted average of the beam-energy-corrected results for the 34 anticollinear measurements taken over the course of this stability measurement, as well as the 6 collinear and 6 anticollinear measurements from the $\nu_0$ determination described in Sec. \ref{sub:bec}, constituting 5580 seconds of total integration time. The uncertainty on $A_{^2P_{1/2}}$ from this analysis is $(0.13)_{\rm stat}(21)_{\rm sys}$ MHz, with the reported statistical uncertainty being the quadrature sum of the propagated error on the weighted mean and the standard error derived from the measurement scatter.
%The estimate for $\nu_0$ was not updated, since the 34 stability measurements were corrected based on the result of the previous anti/collinear $\nu_0$ estimate

% This level of precision will be sufficient to measure many properties of interest across the nuclear chart.
%\textcolor{red}{\cite{TODO}}

\begin{table}
    \centering
    \caption{The rest-frame centroid frequency and $A$ hyperfine coupling constant of $3p\;^2P_{1/2}$ state extracted from fits of the measured spectra. The results are compared with the literature values. The $A$ hyperfine coupling constant of $5s\;^2S_{1/2}$ state was also determined but it is not listed in the table because there is no previous measurement to be compared with. More detailed discussion of hyperfine coupling constants of low-lying states in $^{27}$Al will be given elsewhere\cite{RickeyHyperfine}}
    \begin{ruledtabular}
    \begin{tabular}{ccc}
         & Experiment & Literature\\%ref: DOI:10.1002/andp.19324040406
        \hline
        $\nu_0$ (GHz) & $1129899.838(2)_{\rm stat}(39)_{\rm sys}$ & 1 129 900.176(300) \cite{Martin1979EnergyXIII}  \\
        $A_{^2P_{1/2}}$ (MHz)& $502.93(13)_{\rm stat}(21)_{\rm sys}$ & 502.04(97) \cite{Nak07}\\
        && 503.58(27) \cite{liu08}\\ 
        && 502.0336 (5) \cite{har72}
        % $A_{^2S_{1/2}}$ (MHz)& 136(1) & -
    \end{tabular}
    \end{ruledtabular}
    \label{tab:Constants Table}
\end{table}

\subsection{Systematics}
The uncertainties quoted in TABLE \ref{tab:Constants Table} are dominated by systematic sources. The systematic errors reported are the quadrature sums of the sources enumerated here.
\paragraph{Wavemeter offset}
The fundamental output of the injection-seeded laser is measured with a HighFinesse WSU30 wavemeter, which has an absolute uncertainty of 10~MHz. Since this output is tripled, this yields a 30~MHz uncertainty on the absolute frequency offset. This offset largely cancels for measurements involving frequency differences such as isotope shifts and hyperfine constants ($10^{-5}$~MHz), but remains the dominant source of uncertainty for absolute centroid measurements (30~MHz).

\paragraph{Beam energy}
The anti/collinear analysis described in Sec. \ref{sub:bec} assumes that the beam energy remains constant between the acquisition of the collinear and anticollinear scans. However, it can take up to 1 hour to switch between laser propagation directions, during which time the beam energy could drift. Using the linear fit to the beam energy drift recorded in Sec. \ref{sub:stability}, we estimate a potential residual energy offset of 3.6~mV after the beam energy correction, corresponding to systematic uncertainties of $6\times10^{-5}$~MHz for $A_{^2P_{1/2}}$, and 0.1~MHz for $\nu_0$.

\paragraph{Scan voltage}
The voltage divider used to measure the scanning voltage has a quoted accuracy of 50 parts per million \cite{Dockery2023HyperfineII}. The resulting systematic uncertainty on the hyperfine coefficients was determined by identifying the scanning voltages corresponding to resonance, and scaling them by ($1+50\times10^{-6}$). The resulting beam energies are then converted back to frequencies with the updated Doppler shift, and hyperfine constants are recalculated ($A_{^2P_{1/2}} = (\nu_{F',3}-\nu_{F',2})/3$, for the present case) using the updated resonant frequencies. Taking the difference with the original hyperfine constant value, the resulting systematic uncertainty from this procedure for $A_{^2P_{1/2}}$ is 0.025~MHz.
The systematic error for the centroid can be calculated similarly, using the systematically shifted resonant voltages to instead deduce the centroid
($\nu_0=(5\nu_{2,2'}+7\nu_{3,3'})/12$, for the present case), here contributing 0.03~MHz.
%For cases where the scan range is centered about 0~V, this procedure can be approximated d by simply multiplying the hyperfine constant and (centroid-scan center) each by $5*10^{-5}$, to determine the systematic contribution to the hyperfine constants, and the centroid, respectively.

\paragraph{Beam misalignment}
The ions and resonant excitation laser were aligned through two 5~mm apertures, spaced 2.1~m apart. To be conservative, the maximal deviations were set to 6mm for the ion beam, and 10mm for the laser, to account for post-alignment deviations from beam tuning and laser path drift, respectively. This yields a maximal misalignment angle of $\tan^{-1}\left(\frac{.006}{2.1}\right)+\tan^{-1}\left(\frac{.01}{2.1}\right)\approx$ 8~mrad between the ions and excitation laser. Including this angle into the Doppler shift on resonance results in a potential deviation of 0.02~MHz for $A_{^2P_{1/2}}$. For the absolute centroid, the collinear and anticollinear beam misalignment angles ($\theta_c$ and $\theta_a$) have opposing effects, $\left(\nu_0=\sqrt{\nu_c \nu_a (1+\beta\cos{(\theta_c)})\,(1-\beta\cos{(\theta_a)})}\right)$. Thus, a maximal systematic error of 25~MHz is calculated by assuming 8~mrad of misalignment for one propagation direction, and perfect alignment for the opposite direction. Note that relative measurements, such as isotope shifts, will have a significantly reduced systematic, since the beam misalignment angle is highly reproducible, as can be seen from the $\nu_0$ scatter in FIG \ref{fig:stability}. Here, the alignment procedure was repeated 12 times, yielding centroids consistent to within 3~MHz.

\paragraph{MagneTOF saturation}
When an ion strikes the MagneTOF, a 500~ps pulse is produced, amplified, and recorded by the BECOLA DAQ, capable of discriminating events with 2~ns resolution. Thus, if the instantaneous ion rate on the MagneTOF ever exceeds 1 count per 2~ns, it is possible to saturate the detector, leading to a clipped signal \cite{Simke2024EvaluationBunches}. As the maximum instantaneous count rate recorded during commissioning is 0.03 counts/ns, and the hyperfine spectra are well described by pseudo-Voigt profiles, there is no evidence of clipping in the reported data.
%For future experiments with intense sources and long bunching periods, however, the possibility of detector saturation must be considered.

\paragraph{Random bias}
By comparing measurements for $A_{^2P_{1/2}}$ recorded on different days, a discrepancy in the means was observed that exceeded the statistical uncertainties of the data sets. This observed deviation is too large to be explained by the systematic error sources enumerated above.
A standard approach to account for measurements that differ beyond statistical uncertainties \cite{Lyons1992, Tanabashi2018} is to scale the weighted-mean uncertainty until $\chi_{\rm red}^2 = 1$, which here gives an uncertainty of $0.18$ MHz.
An alternative approach for accounting for measured discrepancy is the method of random systematic bias \cite{Lyons1992}, where each datum is assumed to be biased by some random value, drawn from a normal distribution of mean zero and width $s$. With this assumption, the typical difference between two measurements is $\sqrt{(\sigma_1^2+s^2)+(\sigma_2^2+s^2)}$, where $\sigma_i$ is the statistical uncertainty for $A_{^2P_{1/2}}$ on the $i$th day. By setting this quantity equal to the observed deviation between different sets of measurements, a scale for the systematic random bias can be estimated as $s=\sqrt{\frac{(A_1-A_2)^2-(\sigma_1^2+\sigma_2^2)}{2}} \approx 0.21$~MHz. This approach is reported in TABLE \ref{tab:Constants Table} because it is more conservative, and it is clearer to extrapolate the effect of a random bias to scenarios where multiple measurement periods on the quantity of interest are not possible, e.g. an experiment with rare isotopes.

%\begin{figure}[h]
%\includegraphics[width=0.5\textwidth]{Figures/RandomBiasFigure.png} 
%\caption{\label{fig:bias} \textbf{Random Bias.}
%Best fit estimates for $A_{^2P_{1/2}}$ as extracted from the anti/collinear beam energy analysis dataset (red), and the measurement stability dataset (blue). The dashed lines indicate the statistical-uncertainty-weighted averages for each dataset, and the colored rectangles show one standard error around these averages. 
% For identically distributed measurements, these weighted averages are expected to be consistent on the level of the standard error.
%The deviation observed suggests the presence of a varying random bias.}
%\end{figure}

\subsection{Detector performance}
\begin{figure}[h]
\includegraphics[width=0.5\textwidth]{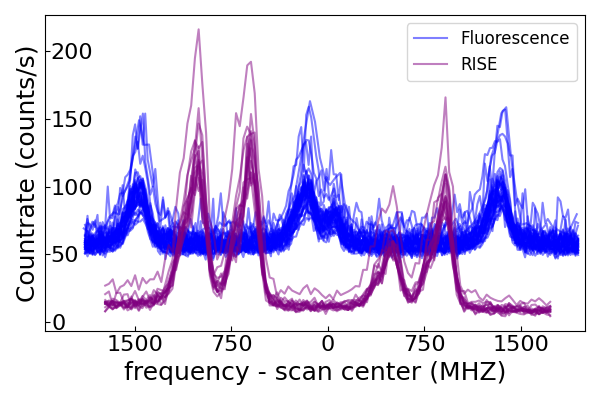}
\caption{\label{fig:comp} \textbf{Comparing Detection Schemes} A collection of spectra for the $(3s^23p)~^2P_{1/2}\rightarrow (3s^24s)~^2S_{1/2}$ transition (blue) as measured with fluorescence collection, as well as for the $(3s^23p)~^2P_{1/2}\rightarrow (3s^25s)~^2S_{1/2}$ transition (purple), as measured with the RISE instrument. To overlap the transitions, all spectra were shifted such that their central scan frequency occurs at 0~MHz. The resonant frequencies differ between the spectra due to the different hyperfine constants for the excited states. The RISE spectra can be seen to exhibit larger signal amplitudes and significantly reduced background count rates.
%The background count rate from fluorescence is roughly 6 times that of resonance ionization, whereas the typical amplitude over background is roughly half that of RISE.
}
\end{figure}

Due to the unknown purity of the ion beam produced by the PIG source, it is not possible to determine an absolute detection efficiency from the $^{27}$Al data. By comparing the current on a Faraday cup located directly after the RFQCB, with the maximum count rate on resonance in a typical spectrum, a lower bound on the absolute efficiency can be set at $\eta\ge10^{-5}$.
Spectra from fluorescence collection were also recorded at BECOLA, using the $(3s^23p)~^2P_{1/2}\rightarrow (3s^24s)~^2S_{1/2}$ transition at 394.4~nm. This $(3s^24s)~^2S_{1/2}$ excited state has a transition probability from the ground state ($A_{ki}=4.99\times10^7s^{-1}$) that is roughly 3.5 times larger than that of the $(3s^25s)~^2S_{1/2}$ state ($A_{ki}=1.42\times10^7s^{-1}$) used for the resonance ionization spectra \cite{Kelleher2008AtomicCompilation}, but shares all angular momentum quantum numbers and as such has equal hyperfine splitting fractions. By running the PIG source and RFQCB at similar settings for measurements on both transitions, RISE's performance can be reasonably characterized by comparing the resulting spectra from each detection scheme. FIG.~\ref{fig:comp} shows a collection of spectra taken for each transition. While the maximum count rates produced from the two schemes are comparable, the average background from fluorescence was observed to be nearly 6~times larger than for resonance ionization, while the typical signal amplitude is roughly half that of resonance ionization. Hence, the signal-to-background ratio was a factor of 10 larger for RISE compared to fluorescence detection for the particular transitions used.

As discussed in Sec. \ref{sub:res}, the signal to background ratio for fluorescence detection is expected to decrease faster than that for resonance ionization when operating with lower yields, as is the case for online experiments on unstable isotopes. It can be concluded from this comparison that the RISE instrument will extend the sensitivity of the BECOLA facility to performing experiments on rare isotopes with significantly lower production rates. Collinear resonance ionization experiments have already demonstrated efficiencies of a few percent, with sensitivities sufficient to measure isotopes produced with yields of just a few ions per second \cite{deGroote2020MeasurementIsotopes}. However, the overall efficiency and signal-to-background ratio depend strongly on the element and beam contaminants.

\section{\label{sec:conclusion}Conclusions and Outlook}
The commissioning of the Resonance Ionization Spectroscopy Experiment (RISE) at BECOLA has demonstrated the feasibility of MHz-level collinear resonance ionization spectroscopy at FRIB. Benchmark measurements on stable $^{27}$Al confirmed the capability of the new instrument to measure high-resolution spectra with long-term stability, and efficient background suppression, establishing RISE as a powerful complement to the existing fluorescence-based collinear laser spectroscopy setup at BECOLA. The ability to perform Doppler-free analyses through collinear and anticollinear configurations further ensures robust extraction of isotope shifts for nuclear charge-radius measurements. RISE is now poised to enable precision spectroscopy of short-lived isotopes produced at FRIB, opening new opportunities to measure changes in the rms nuclear charge radii and nuclear electromagnetic moments of ground and isomeric states across previously inaccessible regions of the nuclear chart. Future developments, including optimized laser schemes, expanded wavelength coverage, and integration with decay spectroscopy, will further enhance the reach of the experiment. The Program Advisory Committee (PAC) at FRIB has already approved a series of experiments with RISE, targeting neutron-deficient aluminum, nickel and zirconium isotopes, neutron-rich silicon and oxygen isotopes, as well as exotic isotopes of francium and thorium. These experiments are in progress and will be carried out in the coming years.
%In addition to the study of atomic spectra to extract nuclear structure properties. Collinear resonance ionization spectrosocpy 

\section{Acknowledgments}
This work was supported by the Office of Nuclear Physics, U.S. Department of Energy, under grants DE-SC0021176, DE-SC0021179, and DE-SC0000661, as well as National Science Foundation, under grant PHY-21-11185. A.B. acknowledges support from the NSF AGEP Fellowship 6949423. We are grateful for the technical support provided by the Bates Laboratory at MIT, and in particular we thank Ernest E. Ihloff for his invaluable assistance during the construction and installation of the RISE instrument. We also extend gratitude to John McGlashing and Joseph D. Cucinotta for their tireless assistance at MIT, as well as for transporting much of the RISE equipment to FRIB. JK acknowledges support from a Feodor-Lynen postdoctoral research fellowship funded by the Alexander-von-Humboldt Foundation. MR acknowledges support from Research Council of Finland grant No. 368322.
% \section{References}

\section{Data Availability}
The data that support the findings of this article are openly available \cite{brinson_2026_18572045}.

\bibliography{references,references2}% Produces the bibliography via BibTeX.

%TODO\\
%Magnetof characterization:\\ %https://www.sciencedirect.com/science/article/pii/S1387380624001489#sec3

\end{document}